\documentclass[letterpaper, 10 pt, journal, twosides]{IEEEtran}

\usepackage[margin=54pt]{geometry}
\usepackage[utf8x]{inputenc}
\usepackage[colorlinks=true, urlcolor=blue, citecolor=blue, linkcolor=blue]{hyperref}
\usepackage{graphics,epsfig,theorem,latexsym,amssymb,amsmath,picinpar,psfrag,subfigure,color}
\usepackage{epstopdf}
\usepackage{dsfont}
\usepackage{mathtools}
\usepackage{mathrsfs, comment}

\def\R{\mathbb R}
\def\RR{\mathbb R}
\def\nats{\mathbb N}
\def\Dp{\mathbb D}

\def\ee{\mathtt{b}}

\newcommand{\tens}{\overline{\otimes}}
\newcommand{\Ltwo}{\mathcal{L}_2((0,1); \RR^{n\chi})}
\newcommand{\Hone}{\mathcal{H}_1((0,1); \RR^{n\chi})}

\def\bea{\begin{eqnarray}}
\def\eea{\end{eqnarray}}
\def\beann{\begin{eqnarray*}}
\def\eeann{\end{eqnarray*}}

\def\cx{{{\mathcal X}}}

\def\ck{{{\mathcal K}}}

\def\be{\begin{equation}}
\def\ee{\end{equation}}
\def\ba{\begin{array}}
\def\ea{\end{array}}
\def\bea{\begin{eqnarray}}
\def\eea{\end{eqnarray}}
\def\beann{\begin{eqnarray*}}
\def\eeann{\end{eqnarray*}}

\newcommand{\Aop}{\mathfrak{A}}

\DeclareMathOperator{\Co}{co}

\DeclareMathOperator{\dom}{dom}

\def\calX{\mathcal X}

\def\A{\mathcal A}

\def\Ltwo{\mathcal{L}_2}
\def\Hone{\mathcal{H}_1}
\def\Dp{\mathbb{D}}

\def\RR{\mathbb R}

\def\ee{\mathtt{b}}

\newtheorem{theorem}{Theorem}
\newtheorem{assumption}{Assumption}
\newtheorem{lemma}{Lemma}
\newtheorem{proposition}{Proposition}
\newtheorem{definition}{Definition}
\newtheorem{remark}{Remark}

\newtheorem{corollary}{Corollary}

\catcode`\@=11

\usepackage{yhmath}
\newcommand{\source}{{THIS IS A PREPRINT VERSION. IF YOU FOUND THIS READING USEFUL FOR YOUR RESEARCH PLEASE CITE THE PUBLISHED VERSION DOI: \href{https://doi.org/10.1109/LCSYS.2021.3081481}{https://doi.org/10.1109/LCSYS.2021.3081481}}}
\usepackage{fancyhdr}
\begin{document}
\def\ps@IEEEtitlepagestyle{}
\title{Synchronization of Identical
Boundary-Actuated Semilinear Infinite-Dimensional Systems\\ (Extended Version)}
\author{Francesco Ferrante, Giacomo Casadei, and Christophe Prieur
\thanks{F. Ferrante \& C. Prieur are with Univ. Grenoble Alpes, CNRS, Grenoble INP, GIPSA-lab 38000 Grenoble, France.} 
\thanks{G. Casadei is with Laboratoire Ampere Dpt. EEA of the \'Ecole Centrale de Lyon, 
Universit\'e de Lyon, 69134 Ecully, France.} \thanks{Research by F. Ferrante and C. Prieur is partially funded by MIAI @Grenoble Alpes (ANR-19-P3IA-0003).}
\thanks{\textcolor{blue}{This files contains fixes to some typos in the published version of the paper. Corrections with respect to the published version are colored in blue and explained in footnotes.}}
}
\maketitle

\begin{abstract}
This paper deals with synchronization of a class of infinite-dimensional systems. The considered network is described by a collection of semilinear Lipschitz boundary-actuated infinite-dimensional dynamics. For undirected connected graphs, sufficient conditions for asymptotic synchronization are established. We show that the proposed conditions when applied to systems of hyperbolic semilinear conservation laws can be recast into a set of matrix inequalities. For this class of systems, sufficient conditions in the form of linear matrix inequalities for the design of synchronizing policies are provided.   
\end{abstract}
\section{Introduction}\label{sec:intro} 
The analysis of collective behaviors has attracted a remarkable attention of the control community due to many applications involving multiple agents interacting over communication networks, e.g, robotic swarms \cite{olfati2004consensus}, power networks  \cite{dorfler2013synchronization}, sensors networks \cite{ferrante2016hybrid}, just to mention a few. In particular, the analysis of consensus and synchronization has seen an increasing interest in the community. Theoretical results on synchronization stem from the seminal work \cite{scardovi2008synchronization} where the case of linear systems was first considered. Nowadays the case of nonlinear agents has been widely studied and a variety of approaches have been proposed to tackle the problem of networks synchronization; see \cite{arcak2007passivity}, \cite{stan2007analysis}, \cite{SWITCH}.

In the last decade, authors have considered the problem of synchronization when the nodes dynamics are described by partial differential equations (\emph{PDEs}) PDEs. One of the first works to consider such a framework is \cite{demetriou2010design} where the author considers the problem of a sensor network observing a process described by a PDE. In \cite{demetriou2013synchronization}, the same author considers the problem of controlling a network of parabolic PDEs and defines a control architecture based on observers to achieve synchronization. For the same class of PDEs, the more challenging case of boundary control has been first considered in \cite{pilloni2015consensus,li2017exact} for first-order parabolic PDEs and recently generalized in \cite{xia2020synchronization} to the case of second-order parabolic PDEs. In \cite{demetriou2018design}, the author provides constructive conditions for the design of control, both at the boundary and in the domain, in networks    
of linear infinite dimensional dynamical systems.
In this letter, we formulate the problem of synchronization for a family of semilinear identical boundary actuated systems and provide conditions under which output synchronization is achieved. The proposed conditions are specialized to a class of hyperbolic semilinear PDEs. Within this setting, we provide constructive conditions for the design of the coupling gain based on the solution to some linear matrix inequalities (\emph{LMI}). 
The contribution with respect to the existing literature on the subject is twofold: i) we consider semilinear infinite dimensional boundary actuated systems; ii) we provide general sufficient conditions and design oriented conditions based on LMIs for the case of semilinear hyperbolic systems. The results obtained are promising and can be extended to other classes of infinite dimensional systems. 

The remainder of the paper is organized as follows. In Section~\ref{sec:Notation}, we introduce some preliminary results that are exploited throughout the paper. Section~\ref{sec:Operator} establishes sufficient conditions for synchronization for a general class of Lipschitz nonlinear infinite-dimensional systems with boundary sensing and actuation. In Section~\ref{sec:Hyperbolic}, the results illustrated in Section~\ref{sec:Operator} are specialized to the case of semilinear hyperbolic PDE. A numerical example is presented in Section~\ref{sec:NumExample}.
\section{Basic Notions and Preliminaries}
\label{sec:Notation}
\subsubsection{Notation}
$\RR$ denotes the set of real numbers,
$\RR_{>0} \coloneqq (0,\infty)$, $\RR^{n}$ is the $n$-dimensional Euclidean space, $\RR^{n\times m}$ is the set of $n\times m$ real matrices, $\Dp_+^{n}$ stands for the set of $n\times n$ positive definite real diagonal matrices, and $\mathds{1}_n\in\RR^n$ is the all-ones vector. Let $X$ be a normed linear space, $x\in X$, and $\mathcal{S}\subset X$ be nonempty, $d(x,\mathcal{S})$ denotes the distance of $x$ to $\mathcal{S}$ and $\vert x\vert$ is the norm of $x$. For a symmetric matrix $M$ negative definiteness and negative semidefiniteness are denoted, respectively as, $M\prec 0$ and $M\preceq 0$.  
Given a symmetric matrix $M\in \RR^{n\times n}$, $\lambda_m(M)$ and $\lambda_M(M)$, denote, respectively, the smallest and the largest eigenvalue of $M$. The symbol $\bullet$ stands for symmetric blocks in symmetric matrices. Let $X$ and $Y$ be normed linear vector spaces, $V\colon X\rightarrow Y$, and $x\in X$, $DV(x)$ is the Fréchet derivative of $V$ at $x$, the symbol $\mathscr{L}(X,Y)$ denotes the set of bounded linear operators from $X$ to $Y$, and $\mathscr{L}(X)\coloneqq\mathscr{L}(X,X)$, and $I_X$ ($I_n$) is the identity operator in $X$ (matrix in $\RR^{n\times n}$). Let $X$ and $Y$ be Hilbert spaces and $T\in\mathscr{L}(X, Y)$, $T^\star$ denotes the adjoint of $T$. Let $\mathscr{U}\subset\R$ be an open interval and $\mathscr{V}\subset\R^n$, $\Ltwo(\mathscr{U}; \mathscr{V})$ denotes the collection of equivalence classes of Lebesgue measurable square integrable functions $f\colon \mathscr{U}\rightarrow \mathscr{V}$ and $\Hone(\mathscr{U};\mathscr{V})$ is the set of $f\in\Ltwo(\mathscr{U}; \mathscr{V})$ such that $f^\prime\in\Ltwo(\mathscr{U}; \mathscr{V})$; where $f^\prime$ stands for the weak derivative of $f$. Let $X$ be a real Hilbert space and $\Aop\colon\dom\Aop\subset X\rightarrow X$ be a linear operator. The notation $\Aop\preceq 0$ indicates that for all $x\in\dom\Aop$, $\langle x, \Aop x\rangle\leq 0$. A selfadjoint operator $P$ on $X$ is called coercive if there exists $\varepsilon>0$ such that for all $x\in X$, $\langle Ax, x\rangle\geq \varepsilon$.
The symbol $\oplus$ stands for the direct sum. Let $X$ and $Y$ be Hilbert spaces, $X\oplus Y$ stands for the Hilbert space $(X\oplus Y, \langle \cdot, \cdot\rangle)$, where $\langle \cdot, \cdot\rangle$ is the ``standard'' inner product on $X\oplus Y$, the equivalent notation $(x, y)=\left[\begin{smallmatrix}x\\y\end{smallmatrix}\right]$ is used for vectors in $X\oplus Y$, and given a positive integer $n$, $X^n\coloneqq\bigoplus_{i=1}^n X$. The notation $\Co\Omega$ indicates the convex hull of the set $\Omega$.
\subsubsection{Graph theory}
A communication graph is described by a triplet ${\mathcal G}=\{{\mathcal V}, {\mathcal E}, A\}$ in which ${\mathcal V}$ is a set of $n$ {\em nodes} ${\mathcal V}=\{v_1,v_2, \ldots, v_n$\}, ${\mathcal E} \subset {\mathcal V}\times {\mathcal V}$ is a set of {\em edges} $e_{jk}$ that models the interconnection between two nodes with the flow of information from node $j$ to node $k$ {\em weighted} by the $(k,j)$-th entry $a_{kj} \geq 0$ of the {\em adjacency matrix} $A \in \mathbb{R}^{n\times n}$.
We denote by $L \in \mathbb{R}^{n \times n}$ the {\em Laplacian matrix} of the graph, with its elements defined as
 $ \ell_{kj} = - a_{kj}$ for $k\ne j$, $\ell_{kj} = \sum_{i=1}^n a_{ki}$ for $k= j$. A graph is said to be connected if and only if $L$ has only one trivial eigenvalue $\lambda_1(L)=0$ and all other eigenvalues $\lambda_2(L), \ldots, \lambda_n(L)$ have strictly positive real part; see \cite{godsil2001algebraic}. We assume the following.
\begin{assumption}\label{assu:connected}\itshape
 Given the graph $\mathcal{G}$, we assume that the associated Laplacian matrix $L$ is connected and diagonalizable. Namely, there exists an orthogonal matrix $T \in \mathbb{R}^{n \times n}$ such that
 \[
\tilde{L}=T^{\top} L T= \begin{bmatrix}
0 & 0_{1 \times (n-1)}\\ 0_{(n-1) \times 1} & \Lambda 
\end{bmatrix}
 \]  
where $\Lambda={\rm diag}(\lambda_2(L),\ldots,\lambda_n(L))$. 
In particular, $T$ can be selected as follows:
 \begin{equation}
 \label{eq:Tcv}
 \begin{aligned}
 & T\coloneqq\begin{bmatrix}
  \frac{1}{\sqrt{n}}\mathds{1}_{n}, M
\end{bmatrix}, \,M^{\top}M=I_{n-1},\, MM^{\top}=I_{n}-\frac{1}{n} \mathds{1}_{n}\mathds{1}_{n}^\top
\end{aligned}
\end{equation}
\hfill$\circ$
\end{assumption} 
\subsubsection{Abstract Dynamical Systems}
In this paper, we consider semilinear abstract dynamical systems of the form:
\begin{equation}\label{eq:AbstractODE}
\dot{x}=\Aop x+f(x)
\end{equation}
where $x\in\cx$ is the system state, $\cx$ is the state space that we assume to be a separable real Hilbert space, $\Aop$ is a linear operator on $\cx$, and $f\colon\cx\rightarrow\cx$ is globally Lipschitz continuous. In particular, similarly as in \cite{magal2018theory}, we consider the following notion of solution to\footnote{In this paper, integrals of Hilbert-space-valued functions are intended in the sense of Bochner.} \eqref{eq:AbstractODE}. 

\begin{definition}
\label{def:solution}
A continuous function $\varphi\colon\dom\varphi\rightarrow \cx$ is a solution to \eqref{eq:AbstractODE} if $\dom\varphi$ is an interval of $\R_{\geq 0}$ including zero and for all $t\in\dom\varphi$
$$
\begin{aligned}
&\int_0^t \varphi(s)ds\in\dom\Aop\\
&\varphi(t)=\varphi(0)+ \Aop \int_0^t \varphi(s)ds+\int_0^t f(\varphi(s))ds
\end{aligned}
$$ 
Moreover, we say that $\varphi$ is maximal if its domain cannot be extended and it is complete if $\sup\dom\varphi=\infty$.
\hfill$\circ$
\end{definition}
\begin{definition}
We say that \eqref{eq:AbstractODE} is well-posed if $\Aop$ is the infinitesimal generator of a strongly continuous semigroup on $\cx$. \hfill$\circ$
\end{definition}
The following result, whose proof is reported in the Appendix, holds.
\begin{proposition}
\label{prop:existence}
Let \eqref{eq:AbstractODE} be well-posed. Then, for any $x_0\in\cx$, there exists a unique maximal solution $\varphi$ to \eqref{eq:AbstractODE} such that $\varphi(0)=x_0$. Moreover, such a solution is complete and it satisfies:
\begin{equation}
\label{eq:SemigroupRep}
\varphi(t)=\mathscr{T}(t)x_0+\int_{0}^t\mathscr{T}(t-s)f(\varphi(s))ds\qquad\forall t\geq 0
\end{equation}
where $\mathscr{T}(t)$ is the strongly continuous semigroup generated by $\Aop$ on $\cx$. In addition, if $x_0\in\dom\Aop$, $\varphi$ is a strong solution to \eqref{eq:AbstractODE}. Namely, $\varphi$ is differentiable almost everywhere, $\dot{\varphi}$ is locally integrable on $\R_{\geq 0}$, and
$\dot{\varphi}(t)=\Aop\varphi(t)+f(\varphi(t))$ almost everywhere. \hfill $\circ$
\end{proposition}
\begin{definition}
Let $\mathcal{S}\subset\cx$ be closed. We say that $\mathcal{S}$ is globally exponentially stable for  \eqref{eq:AbstractODE} if any maximal solution $\varphi$ to \eqref{eq:AbstractODE} is complete and it satisfies:
$$
d(\varphi(t), {\mathcal{S}})\leq \kappa e^{-\lambda t} d(\varphi(0), \mathcal{S})\qquad\forall t\geq 0
$$ 
for some (solution independent) $\lambda, \kappa>0$.\hfill$\circ$
\end{definition}
The result given next, whose proof is reported in the Appendix, provides sufficient conditions for stability of a closed set.
\begin{theorem}
\label{thm:GES}
Let \eqref{eq:AbstractODE} be well-posed and $\mathcal{S}\subset\cx$ be closed. Assume that there exists a Fréchet differentiable functional $V\colon\cx\rightarrow\R$ and positive scalars $\alpha_1, \alpha_2, \alpha_3$, and $p$ such that the following items hold:
\begin{itemize}
\item[($i$)] For all $x\in\cx$, 
$
\alpha_1 d(x,\mathcal{S})\leq V(x)\leq\alpha_2 d(x,\mathcal{S})
$ 
\item[($ii$)] For all $x\in\dom\Aop$, 
$
DV(x)(\Aop x+f(x))\leq-\alpha_3 d(x,\mathcal{S})
$ 
\end{itemize}

Then, $\mathcal{S}$ is globally exponentially stable for \eqref{eq:AbstractODE}.\hfill$\circ$
\end{theorem}
\subsubsection{Kronecker product for linear operators and properties}
In this paper, we make use of the following extension of the Kronecker product\footnote{The extension of the Kronecker product in Definition~\ref{def:kron} is nothing but the tensor product of the operators $S$ and $\Pi$ if one identifies $\R^{\theta}\otimes V\cong V^\theta$, with $V^\theta\coloneqq\bigoplus_{i=1}^\theta V$.}.  
\begin{definition}
\label{def:kron}
Let $S\in\R^{\theta\times\theta}$, $U$, $V$ be linear spaces, and $\Pi\colon U\rightarrow V$ be a linear operator. We define:
$$
\begin{aligned}
&S \otimes \Pi\colon\bigoplus_{i=1}^\theta U\rightarrow\bigoplus_{i=1}^\theta V, \,
&S\otimes \Pi\coloneqq\begin{bmatrix}
s_{1,1}\Pi&\dots&s_{1,\theta}\Pi\\
s_{2,1}\Pi&\dots&s_{2,\theta}\Pi\\
\vdots&\dots&s_{\theta,\theta}\Pi
\end{bmatrix}
\end{aligned}
$$
\hfill$\circ$
\end{definition}

In the following we use the symbol $\otimes$ to indifferently represent the Kronecker product in the classical or in the operator sense. 
\section{Synchronization of Abstract Dynamical Systems with Boundary Inputs}\label{sec:Operator}
We consider a network of $n$ identical boundary-actuated systems formally written as:
\begin{equation}
\begin{aligned}
\dot{x}_i&=\mathfrak{A}x_i+\mathfrak{E}f(q_i)&\\
\mathfrak{B}x_i&=Bu_i\\
q_i&=\mathfrak{Q}x_i\\
y_i&=\mathfrak{C}x_i&i=1, 2, \dots, n
\end{aligned}
\label{eq:plant}
\end{equation}
where $x_i\in\mathcal{X}$ is the plant state, $u_i\in\mathbb{R}^{n_u}$ the plant input, and $y_i\in\mathbb{R}^{n_y}$ is the plant output. The state space $\calX$ is a real separable Hilbert space. 
The system is defined by the following operators: $\mathfrak{A}\colon\dom\mathfrak{A}\rightarrow\mathcal{X}$, $\mathfrak{B}\colon\dom\mathfrak{B}\rightarrow\mathcal{U}$, $\mathfrak{C}\colon\dom\mathfrak{C}\rightarrow\mathbb{R}^{n_y}$,  $\mathfrak{Q}\in\mathscr{L}(\mathcal{X},\mathcal{Q})$, $\mathfrak{E}\in\mathscr{L}(\mathcal{E}, \mathcal{X})$,
and $B\colon\mathbb{R}^{n_u}\rightarrow\mathcal{U}$ where $\mathcal{U}$ is a real vector space, $\mathcal{Q}$ and $\mathcal{E}$ are real Hilbert spaces,
$\dom\mathfrak{A}\subset\dom\mathfrak{C}\cap\dom\mathfrak{B}$,
and
$f\colon\mathcal{Q}\rightarrow\mathcal{E}$. The following assumption is considered throughout the paper.

\begin{assumption}\label{assu:Incremental}
The function $f\colon\mathcal{Q}\rightarrow\mathcal{E}$ satisfies the following conditions:
\begin{itemize}
\item $f$ is globally Lipschitz continuous on $\mathcal{Q}$. 
\item there exists $\mathcal{K}\in\mathscr{L}(\mathcal{Q}\oplus\mathcal{E})$ selfadjoint such that for all $\zeta_1, \zeta_2\in\mathcal{Q}$:
\begin{equation}
\scalebox{0.95}{$\begin{array}{l}
\left\langle \begin{bmatrix}
\zeta_1-\zeta_2\\
f(\zeta_1)-f(\zeta_2)
\end{bmatrix},\underbrace{\begin{bmatrix}
\ck_{11} & \ck_{12}\\ \ck_{12}^\star & \ck_{22}
\end{bmatrix}}_\mathcal{K}\begin{bmatrix}
\zeta_1-\zeta_2\\
f(\zeta_1)-f(\zeta_2)
\end{bmatrix}\right\rangle\geq 0
\end{array}$}
\label{eq:QadraticIneq}
\end{equation}
\end{itemize}
\hfill$\circ$
\end{assumption}
 \begin{remark}
 Assumption~\ref{assu:Incremental} is a natural extension to the infinite dimensional case of well known incremental properties; see, e.g, \cite{zhang2014fully} and \cite{scardovi2010synchronization} just to mention a few notable examples.
 \end{remark}
 
Systems \eqref{eq:plant} exchange their outputs according to a graph $\mathcal{G}$ and the input is designed as a diffusive coupling of the form
\begin{equation}\label{eq:controlLap}
u_i=K \sum_{j=1}^n \ell_{ij} \mathfrak{C}x_j 
\end{equation}
where $\ell_{ij}$ are elements of the Laplacian matrix $L$ associated to $\mathcal{G}$. 
The choice of \eqref{eq:controlLap} is ubiquitous in the finite dimensional case when agents are, as in our case, identical. 

By defining $\mathbf{x}\coloneqq(x_1, x_2, \dots, x_n)$, 
the network of \eqref{eq:plant} interconnected through \eqref{eq:controlLap} can be written as the following abstract dynamical system: 
\begin{subequations}
\label{eq:closed-loop_final}
\begin{equation}
\dot{\mathbf{x}}=\mathcal{A}\mathbf{x}+(I_n \otimes \mathfrak{E})F((I_n \otimes \mathfrak{Q})\mathbf{x})
\end{equation}
where:
\begin{equation}
\label{eq:opA}
\begin{aligned}
&\mathcal{A}\coloneqq I_n \otimes \mathfrak{A}\\
&\dom\mathcal{A}\coloneqq\left(\bigoplus_{i=1}^n\dom\mathfrak{A}\right)\cap\ker\left((I_n \otimes \mathfrak{B})-(L \otimes BK\mathfrak{C})\right)\end{aligned}
\end{equation}
\end{subequations}
and\footnote{\textcolor{blue}{The definition of $F$ is missing.}} 
\textcolor{blue}{$\mathbf{x} \mapsto F(\mathbf{x})\coloneqq (f(q_1), f(q_2), \dots, f(q_n))$, where for all $i=1, 2,\dots, n$, $q_i$ is defined in \eqref{eq:plant}.}

The following assumption is considered in the paper.
\begin{assumption}
\label{assu:WellPosed}
System \eqref{eq:closed-loop_final} is well posed.\hfill$\circ$
\end{assumption}
\begin{proposition}\label{prop:MR}
Let Assumption~\ref{assu:connected}, Assumption~\ref{assu:Incremental}, and Assumption~\ref{assu:WellPosed} hold. Suppose that there exist $\alpha, \tau>0$, and $P\in\mathscr{L}(\calX, \calX)$ selfadjoint and coercive such that for all $i\in\{2, 3,\dots, n\}$
\begin{eqnarray}
&& \hspace{-.5cm} \Upsilon_i\coloneqq\left[
\begin{array}{c|c}
   P \mathfrak{A}_{i}+\frac{\tau}{2}  \mathfrak{Q}^\star \mathcal{K}_{11} \mathfrak{Q}+\frac{\alpha}{2} P &  P\mathfrak{E}+\frac{\tau}{2} \mathfrak{Q}^\star \mathcal{K}_{12}\\[.6em] 
  \hline
\frac{\tau}{2}\mathcal{K}_{12}^\star\mathfrak{Q}& \frac{\tau}{2} \mathcal{K}_{22} \\
\end{array}
\right]\preceq 0 \nonumber \\
&&\hspace{-.5cm} \dom\Upsilon_i\coloneqq \dom\mathfrak{A}_{i}\oplus\mathcal{E}
\label{eq:OP_ineq}\\
&&\hspace{-.5cm} \mathfrak{A}_{i}\coloneqq\mathfrak{A}\qquad\dom\mathfrak{A}_{i}\coloneqq\dom\mathfrak{A}\cap\ker
\left(\mathfrak{B}-\lambda_i BK\mathfrak{C}\right) 
\label{eq:Afraklambda}
\end{eqnarray}
where $\mathcal{K}_{ij}$'s are defined in \eqref{eq:QadraticIneq} and $\lambda_i$ are the nonzero eigenvalues of the Laplacian $L$, according to Assumption~\ref{assu:connected}. Then, the closed set
\begin{equation}
\label{bigw}
\mathcal{S}\coloneqq\{\mathbf{x}\in \mathcal{X}^n\colon x_1=x_2=\cdots= x_{n}\}
\end{equation}
is globally exponentially stable for \eqref{eq:closed-loop_final}. 
\end{proposition}
\begin{IEEEproof}
Let us define the following change of variables:
\begin{equation}\label{eq:coctot}
{\bf x} \mapsto {\bf w}=(w_1, w_2,\dots, w_n)\coloneqq (T^{\top} \otimes I_{\calX}) {\bf x}
\end{equation}
where $T$ is selected as in \eqref{eq:Tcv}. By construction of $T$, it follows that ${\bf x}\in\mathcal{S}$ if and only if
for all $i=2,\ldots,n$, $w_i=0$. In particular, $d({\bf w}, \mathcal{S})=\vert {\bf e}\vert$, with ${\bf e}\coloneqq (w_2, \ldots, w_n)\in{\cal X}^{n-1}$.
Focusing on the synchronization error dynamics ${\bf e}$, one gets:
\begin{equation}
\dot{\mathbf{e}}=\mathfrak{A}_e\mathbf{e}+(M^{\top} \otimes \mathfrak{E})F((T \otimes \mathfrak{Q})\mathbf{w})
\label{eq:PDEerror}
\end{equation}
where $\mathfrak{A}_e \coloneqq \bigoplus_{i=2}^n \mathfrak{A}_i$, $\dom\mathfrak{A}_{e}\coloneqq \bigoplus_{i=1}^n\dom \mathfrak{A}_i$.
Bearing in mind \eqref{eq:PDEerror}, now we analyze the dynamics of $\mathbf{w}$ via the following Lyapunov candidate:
\begin{equation}
\label{eq:VinpropFunctional}
V({\bf w})\coloneqq \langle \mathbf{P} {\bf e}, {\bf e}\rangle\quad\forall {\bf w}\in\calX^n
\end{equation}
with $\mathbf{P}\coloneqq I_{n-1}\otimes P$. Notice that since $P$ is a bounded coercive operator, there exist $\alpha_1, \alpha_2>0$ such that for all ${\bf w}\in\calX^n$, $\alpha_1 d({\bf w}, \mathcal{S})^2=\alpha_1 \vert{\bf e}\vert^2 \leq V({\bf w})\leq \alpha_2 \vert{\bf e}\vert^2=\alpha_2 d({\bf w}, \mathcal{S})^2$. Thus item $(i)$ of Theorem~\ref{thm:GES} holds for $V$. We now show that item $(ii)$ of Theorem~\ref{thm:GES} holds.
From direct computation of $DV$, with a slight abuse of notation, one has that for all ${\bf w}\in\mathcal{O}\coloneqq(\dom\mathfrak{A}\cap\ker\mathfrak{B})\oplus \dom\mathfrak{A}_{\bf e}$:
\begin{align*}
\dot{V}({\bf w})&\coloneqq DV({\bf e})
\left(\mathfrak{A}_{\bf e}{\bf e}+(M^\top \otimes \mathfrak{E})F((T \otimes \mathfrak{Q})\mathbf{w})\right)\\
&=2\langle {\bf P e}, \mathfrak{A}_{\bf e}{\bf e}+(M^\top \otimes \mathfrak{E})F((T \otimes \mathfrak{Q})\mathbf{w}) \rangle.
\end{align*} 
By defining\footnote{\textcolor{blue}{$\mathfrak{E}$ should be $\mathcal{E}$. Same below.}} $\Delta F \coloneqq(M^\top \otimes I_{\textcolor{blue}{\mathcal{E}}}) F((T \otimes \mathfrak{Q})\mathbf{w})$
and $\xi\coloneqq({\bf e}, \Delta F)$, one has $\Gamma({\bf w})\coloneqq\dot{V}({\bf w})+\alpha V(\mathbf{w})=2\langle \xi, \Theta \xi \rangle$, with
$$
\Theta\coloneqq\left[\begin{array}{c|c}
  (I_{n-1} \otimes P)(\mathfrak{A}_{e}+\frac{\alpha}{2}I_{\mathcal{X}^{n-1}}) & I_{n-1} \otimes (P\mathfrak{E})  \\[.6em]
  \hline
0& 0\\
\end{array}\right].
$$
Note that, by rewriting $\xi$ as 
\begin{align*}
\xi &= \begin{bmatrix}
(M^\top \otimes I_\calX) & 0\\
0 & (M^\top \otimes I_{\textcolor{blue}{\mathcal{E}}})\end{bmatrix}\begin{bmatrix}
\mathbf{x}\\
F((I_n \otimes \mathfrak{Q}))\mathbf{x})
\end{bmatrix}
\end{align*}
and by setting 
\begin{equation}
\mathfrak{K}\coloneqq\left[
\begin{array}{c|c}
   I_{n-1} \otimes \mathfrak{Q}^\star \mathcal{K}_{11} \mathfrak{Q} & I_{n-1} \otimes  \mathfrak{Q}^\star \mathcal{K}_{12}  \\[.6em]
  \hline
 I_{n-1} \otimes \mathcal{K}_{12}^\star \mathfrak{Q}  & I_{n-1} \otimes \mathcal{K}_{22} \\
\end{array}
\right]
\label{eq:Kfrak}
\end{equation}
in view of Lemma~\ref{appendix:lemmino}, we have that
\begin{align*}
\langle \xi,\mathfrak{K} \xi\rangle=&\frac{1}{n} \sum_{1\leq i \leq j \leq n}\left\langle
\begin{bmatrix}
\mathfrak{Q}x_i-\mathfrak{Q}x_j\\
f(\mathfrak{Q}x_i)-f(\mathfrak{Q}x_j)
\end{bmatrix},\right.\\
&\mathcal{K} \left.\begin{bmatrix}
\mathfrak{Q} x_i- \mathfrak{Q} x_j\\
f(\mathfrak{Q}x_i)-f(\mathfrak{Q}x_j)
\end{bmatrix}\right\rangle \end{align*}
Hence, using \eqref{eq:QadraticIneq}, for all $\xi\in\calX^{n-1}\oplus\mathcal{E}^{n-1}$, $\langle \xi,\mathfrak{K} \xi\rangle\geq 0$.
Then, it is readily seen that for all ${\bf w}\in\mathcal{O}$
\begin{equation}
\label{eq:BoundVdot}
\begin{aligned}
\Gamma({\bf w})&\leq 2\langle \xi, \Theta \xi \rangle+ \tau\langle \xi,\mathfrak{K} \xi\rangle=2\left\langle \xi , \left(\Theta+\frac{\tau}{2}\mathfrak{K}\right)\xi \right\rangle
\end{aligned}
\end{equation}
At this stage, observe that\footnote{\textcolor{blue}{Missing $\mathfrak{Q}$'s in the equation below (15).}}:
\begin{align*}
\Theta+\frac{\tau}{2}\mathfrak{K}&=\left[
\begin{array}{c|c}
   (I_{n-1} \otimes P)(\mathfrak{A}_{e}+\frac{\alpha}{2}I_{\mathcal{X}^{n-1}}) & (I_{n-1} \otimes P\mathfrak{E})  \\[.6em]
  \hline
 0 & 0 \\
\end{array}
\right]\\ &\hspace*{-0.1cm}
+\frac{\tau}{2}\left[
\textcolor{blue}{\begin{array}{c|c}
  (I_{n-1} \otimes (\mathfrak{Q}^\star\mathcal{K}_{11}\mathfrak{Q})) & (I_{n-1} \otimes (\mathfrak{Q}^\star\mathcal{K}_{12})  \\[.6em]
  \hline
 (I_{n-1} \otimes \mathcal{K}_{12}^\star\mathfrak{Q})  & (I_{n-1} \otimes \mathcal{K}_{22}) \\
\end{array}}
\right]
\end{align*}
which, due to the block diagonal structure of each sub-partition of the operator, by opportunely reordering of the blocks can be seen as a block diagonal operator with blocks $\Upsilon_i$ as defined in \eqref{eq:OP_ineq}.  Therefore, combining \eqref{eq:OP_ineq} and \eqref{eq:BoundVdot}, it follows that for all ${\bf w}\in\mathcal{O}$,
$
\Gamma({\bf w})=\dot{V}({\bf w})+\alpha V({\bf w})\leq 0
$.
Thus, by invoking Theorem~\ref{thm:GES} the result is established.
\end{IEEEproof}

\begin{remark}
It is worth pointing out that, similarly to the finite-dimensional case, the topology, and in particular the eigenvalues $\lambda_i$, of the Laplacian $L$ play a structural role in ensuring  convergence towards synchronization, this is highlighted by \eqref{eq:Afraklambda}. 
\end{remark}

\section{Application to hyperbolic systems with incrementally cone-bounded nonlinearities}
\label{sec:Hyperbolic}
We now illustrate the applicability of our results in the specific case of a semilinear system of hyperbolic PDEs. In particular, we consider a collection of $n$ identical systems formally represented as:
\begin{equation}
\begin{aligned}
&\frac{\partial}{\partial t} x_i(t, \chi)+S\frac{\partial}{\partial \chi} x_i(t, \chi)+E h(z_i(t, \chi))=0\\
&z_i(t, \chi)=Qx_i(t, \chi)\\
&x_i(t, 0)=Hx_i(t, 1)+Bu_i(t)&\!\!\!\!\!\!\!\!\!\!\!\!\!\!\!\!\!\!\forall i=1,2,\dots, n
\end{aligned}
\label{eq:hyperbolic}
\end{equation}
where $\chi\in(0, 1)$, $t>0$ are, respectively, the spatial and temporal variables, $x_i\in\R^{n_p}$, $z_i\in\R^{n_z}$, 
 $H\in\R^{n_p\times n_p}$, $B\in\R^{n_p\times n_u}$, $S\in\Dp_+^{n_p}$, $E\in\R^{n_p\times n_e}$, $Q\in\R^{n_z\times n_p}$, and $h\in\R^{n_z}\rightarrow\R^{n_e}$ is incrementally cone-bounded, i.e., for all 
$q_1, q_2\in\R^{n_z}$
\begin{equation}
\begin{bmatrix}
q_1-q_2\\
h(q_1)-h(q_2)
\end{bmatrix}^\top\begin{bmatrix}
0&G_{12}\\
G_{12}^\top&G_{22}
\end{bmatrix} \begin{bmatrix}
q_1-q_2\\
h(q_1)-h(q_2)
\end{bmatrix}\geq 0
\label{eq:QuadraticHyperbolic}
\end{equation}
where $G_{12}$ and $G_{22}\prec 0$ are some given matrices. Notice that the satisfaction of \eqref{eq:QuadraticHyperbolic} implies that $h$ is globally Lipschitz continuous\footnote{Lipschitz continuity of $h$ can be shown by observing that
\eqref{eq:QuadraticHyperbolic} implies:
$$
\begin{aligned}
&2 (q_1-q_2)^\top G_{12}(h(q_1)-h(q_2))+\\
&(h(q_1)-h(q_2))^\top G_{22}(h(q_1)-h(q_2))\geq 0
\end{aligned}
$$
which, combined to Young's inequality, yields 
$$
\begin{aligned}
&0\leq 2 (q_1-q_2)^\top G_{12}(h(q_1)-h(q_2))+\\
&(h(q_1)-h(q_2))^\top G_{22}(h(q_1)-h(q_2))\leq\\
&\frac{1}{\varpi}(q_1-q_2)^\top G_{12} G_{12}^\top(q_1-q_2)+\\
&(h(q_1)-h(q_2))^\top (\varpi I+G_{22})(h(q_1)-h(q_2))
\end{aligned}
$$
By taking $\varpi$ small enough so that $\varpi I+G_{22}\prec 0$, the above inequality implies that $h$ is globally Lipschitz.
}.
\begin{remark}
It is worth to mention that incremental sector bounded nonlinearities as in \cite{zhang2014fully} can be transformed into incremental cone bounded nonlinearities as in \eqref{eq:QuadraticHyperbolic} via a simple change of variables.
\end{remark}
In terms of available information, we assume that only the state at the boundary $\chi=1$ is available for communication.
In this setting, by taking as a state space $\calX\coloneqq\Ltwo((0,1);\R^{n_p})$ (endowed with the standard inner product), system $i$ can be represented as \eqref{eq:plant}
with:
\begin{equation}
\begin{aligned}
&\dom\mathfrak{A}=\Hone((0,1); \R^{n_p})&&\mathfrak{A}\coloneqq -S\frac{d}{d\chi}\\
&\dom\mathfrak{B}=\Hone((0,1); \R^{n_p})&&\mathfrak{B}r\coloneqq r(0)-Hr(1)\\
&\dom\mathfrak{C}=\Hone((0,1); \R^{n_p})&&\mathfrak{C}r\coloneqq r(1)\\
&\dom\mathfrak{Q}=\Ltwo((0,1); \R^{n_z})&&(\mathfrak{Q}r)(\chi)\coloneqq Qr(\chi)\\
&\dom\mathfrak{E}=\Ltwo((0,1); \R^{n_e})&&(\mathfrak{E}g)(\chi)\coloneqq -Eg(\chi)
\end{aligned}\label{eq:plant_hyperbolic} 
\end{equation}
and $f\colon\Ltwo((0,1); \R^{n_z})\rightarrow\Ltwo((0,1); \R^{n_e})$ defined as
$
(f(s))(\chi)=h(s(\chi))
$, which turns out to be globally Lipschitz continuous due to $h$ being so. In particular, by taking $\dom\mathcal{K}=\Ltwo((0,1); \R^{n_e+n_z})$ and
$$
\begin{array}{ll}
(\mathcal{K}g)(\chi)=\begin{bmatrix}
0&G_{12}\\
G_{12}^\top &G_{22}
\end{bmatrix}g(\chi)
\end{array}
$$
from \eqref{eq:QuadraticHyperbolic}, it turns out that for all $\zeta_1, \zeta_2\in\Ltwo((0,1);\R^{n_z})$
$$
\left\langle \begin{bmatrix}
\zeta_1-\zeta_2\\
f(\zeta_1)-f(\zeta_2)
\end{bmatrix}, \mathcal{K}\begin{bmatrix}
\zeta_1-\zeta_2\\
f(\zeta_1)-f(\zeta_2)
\end{bmatrix}\right\rangle\geq 0
$$
The latter, together with the Lipschitzness of $f$, ensures that  
Assumption~\ref{assu:Incremental} holds. The family of systems \eqref{eq:plant_hyperbolic} is interconnected over a graph $\mathcal{G}$ through the inputs $u_i$, for all $i\in\{1, 2,\dots, n\}$, defined in \eqref{eq:controlLap}. It is worth to remark that the interconnected system is a system of $n$ 1-D semilinear balance laws and for this specific case Assumption~\ref{assu:WellPosed} holds; see, e.g., \cite{bastin2016stability}. In the following section, we provide sufficient  conditions under which this network achieves synchronization. As a second step, we give constructive conditions for the design of the coupling gain $K$ ensuring synchronization.
\subsection{Sufficient conditions for synchronization}
The result given next provides sufficient conditions to ensure synchronization of the network of \eqref{eq:hyperbolic}. In the remainder of the paper, we use the following notation $
\underline{\lambda}\coloneqq\min\{\lambda_2(L),\lambda_3(L),\dots,\lambda_n(L)\}$ and $\overline{\lambda}\coloneqq\max\{\lambda_2(L),\lambda_3(L),\dots,\lambda_n(L)\}
$.
\begin{proposition}
\label{prop:AnalysisHyperbolic}
Let $K\in\R^{n_u\times n_p}$ be given. If there exist $R\in\mathbb{D}^{n_p}_+$, 
and positive scalars $\mu, \alpha$, and $\tau$ such that
\begin{align}
\label{eq:HyperbolicAssu1}
&\begin{bmatrix}
-e^{-\mu}R&(H+\varrho BK)^\top R\\
\bullet&-R
\end{bmatrix}\preceq 0&\varrho\in\{\underline{\lambda}, \overline{\lambda}\}\\[0.5cm]
&\Phi(\chi)\prec 0&\forall \chi\in\{0, 1\}
\label{eq:HyperbolicAssu2}
\end{align}
where for all $\chi\in[0, 1]$
\begin{equation}
\label{eq:Qchi}
\Phi(\chi)\coloneqq\begin{bmatrix}
-\mu e^{-\mu\chi}R&-e^{-\mu\chi}S^{-1}RE+\tau Q^\top G_{12}\\
\bullet&\tau G_{22}
\end{bmatrix}
\end{equation}
Then, the set  $\mathcal{S}$ in \eqref{bigw} is globally exponentially stable for the family of systems \eqref{eq:hyperbolic} interconnected via \eqref{eq:controlLap}.
\end{proposition}
\begin{IEEEproof}
The proof is organized in two steps.

{\noindent\textit Step 1:} We show that \eqref{eq:HyperbolicAssu1} is equivalent to:
\begin{equation}
\label{eq:Tneg}
T(\lambda)\coloneqq (H+\lambda BK)^\top  R(H+\lambda BK)-
e^{-\mu}R\preceq 0 \,
\end{equation}
for $\lambda\in [\underline{\lambda},\overline{\lambda}]$. To this end, observe that thanks to Schur complement and a simple congruence transformation, $T(\lambda)\preceq 0$ if and only if
\begin{equation}
\Omega(\lambda)\coloneqq\begin{bmatrix}
-e^{-\mu}R&(H+\lambda BK)^\top R\\
\bullet&-R
\end{bmatrix}\preceq 0
\label{eq:Mla}
\end{equation}
Now observe that there exist $\gamma_1,\gamma_2\colon[\underline{\lambda},\overline{\lambda}]\rightarrow [0, 1]$ such that $\lambda=\underline{\lambda}\gamma_1(\lambda)+\overline{\lambda}\gamma_2(\lambda)$ and $\gamma_1(\lambda)+\gamma_2(\lambda)=1$.  Hence, due to the affine dependence of $\Omega$ on $\lambda$, one has
$
\Omega([\underline{\lambda}, \overline{\lambda}])=\Co\{\Omega(\underline{\lambda}), \Omega(\overline{\lambda})\}
$. This shows that \eqref{eq:Tneg} is equivalent to \eqref{eq:HyperbolicAssu1}.

{\textit Step 2:} Let $P\in\mathscr{L}(\Ltwo((0,1);\R^{n_p}))$ defined as:
$$
(Ph)(\chi)\coloneqq e^{-\mu \chi}S^{-1}Rh(\chi)\qquad\forall \chi\in(0, 1)
$$
which, due to $R,S\in\mathbb{D}^{n_p}_+$, is a coercive operator. 
Then, for all $h\in\dom\mathfrak{A}$ one has:
$$
(P\mathfrak{A}h)(\chi)=-e^{-\mu \chi} R\frac{d}{d\chi}h(\chi)\qquad\forall \chi\in(0, 1)
$$
Pick $i\in\{2,\dots, n\}$ and let $\Upsilon_i$ be defined as in \eqref{eq:OP_ineq} with $\alpha>0$ to be selected later. Observe that, due to the above selection of $P$ and \eqref{eq:plant_hyperbolic}, one has that, for all $\zeta\coloneqq(\zeta_1,\zeta_2)\in\dom\Upsilon_i$
$$
\begin{aligned}
\langle \zeta,\Upsilon_i\zeta\rangle=&-\int_0^1 e^{-\mu \chi}\zeta_1^\top(\chi)R\left(  \frac{d}{d\chi}-\frac{\alpha}{2}I\right)\zeta_1(\chi) d\chi\\
&+\int_0^1 \zeta_1^\top(\chi) (-e^{-\mu z}S^{-1} R E+\tau G_{12}Q) \zeta_2(\chi)d\chi\\
+&\frac{\tau}{2} \int_0^1 \zeta_2^\top(\chi) G_{22} \zeta_2(\chi)d\chi
\end{aligned}
$$
By integrating by parts the first term in the above expression and using the fact that $\zeta\in\dom\Upsilon_{i}$, one has:
\begin{equation}
\begin{aligned}
\langle \zeta,\Upsilon_i \zeta\rangle=&\frac{1}{2}\zeta_1(1)^\top T(\lambda_i)\zeta_1(1)
+\frac{1}{2}\int_0^1\!\!
\zeta(\chi)^\top \Phi^\prime(\chi)
\zeta(\chi)d\chi\\
\end{aligned}
\label{eq:finalBound}
\end{equation}
where $\Phi^\prime(\chi)\coloneqq\Phi(\chi)+\alpha e^{-\mu\chi}R$ and $T(\lambda_i)$ and $\Phi(\chi)$ are defined, respectively, 
 in \eqref{eq:Tneg} and \eqref{eq:Qchi}. 
To conclude, notice that due to the affine dependence of $\Phi$ on $e^{-\mu\chi}$, one has $\Phi([0, 1])=\Co\{\Phi(0), \Phi(1)\}$. Hence, by using \eqref{eq:HyperbolicAssu2}, for all $\chi\in[0, 1]$, $\Phi(\chi)\prec 0$, which by taking $\alpha$ small enough gives, for all $\chi\in[0, 1]$, $\Phi^\prime(\chi)\prec 0$. Combining this with \eqref{eq:Tneg}, \eqref{eq:finalBound} yields $\langle \zeta,\Upsilon_i\zeta\rangle\leq 0$.
To summarize, the above analysis shows that there exists $\alpha>0$ such that for all $i\in\{2,\dots, n\}$ and $\zeta\in\dom\Upsilon_i$, $\langle \zeta,\Upsilon_i\zeta\rangle\leq 0$.
Invoking Proposition~\ref{prop:MR} establishes the result.
\end{IEEEproof}

\begin{remark}
Compared to Proposition~\ref{prop:MR}, condition \eqref{eq:HyperbolicAssu1} is to be checked only for the Laplacian's eigenvalues $\underline{\lambda}$ and $\overline{\lambda}$. This is reminiscent of the results on synchronization of linear discrete-time finite-dimensional systems (see, e.g., \cite{you2011network}), where necessary and sufficient conditions for synchronization depend $\underline{\lambda}$ and $\overline{\lambda}$. In contrast with this, condition \eqref{eq:HyperbolicAssu2} is \textit{only} sufficient and allows us to design a common $K$ as detailed in the next section.
\end{remark}

\subsection{Sufficient conditions for the design of the coupling gain}
Proposition~\ref{prop:AnalysisHyperbolic} establishes sufficient conditions in the form of matrix inequalities to check if a given coupling gain $K$ ensures asymptotic synchronization. However, such a result is difficult to exploit directly for design purposes. 
To overcome this problem, next we provide a result particularly suited for design.
\begin{corollary}
\label{coro:design}
If there exists $W\in\Dp_+^{n_p}$, $Y\in\R^{n_u\times n_p}$, and $\sigma>0$ such that
\begin{align}
\label{eq:HyperbolicAssu1:design}
&\begin{bmatrix}
-e^{-\mu}W&WH^\top+\varrho Y^\top B^\top\\
\bullet&-W
\end{bmatrix}\preceq 0&\varrho\in\{\underline{\lambda}, \overline{\lambda}\}\\[0.5cm]
&\Phi_s(\chi)\prec 0&\forall \chi\in\{0, 1\}
\label{eq:HyperbolicAssu2:design}
\end{align}
where for all $\chi\in[0, 1]$
$$
\Phi_s(\chi)\coloneqq\begin{bmatrix}
-\mu e^{-\mu\chi}W&-\sigma e^{-\mu\chi}S^{-1}E+WQ^\top G_{12}\\
\bullet&\sigma G_{22}
\end{bmatrix}
$$
Then, $K=YW^{-1}$ ensures that the set \eqref{bigw} is globally exponentially stable for the family of systems \eqref{eq:hyperbolic} interconnected via \eqref{eq:controlLap}.
\end{corollary}
\begin{IEEEproof}
The proof hinges upon Proposition~\ref{prop:AnalysisHyperbolic}. In particular, we show that the satisfaction of \eqref{eq:HyperbolicAssu1:design} and \eqref{eq:HyperbolicAssu2:design} is equivalent to \eqref{eq:HyperbolicAssu1} and \eqref{eq:HyperbolicAssu2} with $W=R^{-1}$, $\sigma^{-1}=\tau$, and $K=YW^{-1}$. By pre-and-post multiplying the matrix in the lefthand side of  \eqref{eq:HyperbolicAssu1:design} by $W^{-1}\oplus W^{-1}$, one gets the matrix in the lefthand side of \eqref{eq:HyperbolicAssu1} with $R=W^{-1}$. This shows that  \eqref{eq:HyperbolicAssu1:design} is equivalent to \eqref{eq:HyperbolicAssu1}. To conclude, bearing in mind that $S$ and $W$ are diagonal, by pre-and-post multiplying $\Phi_s(\chi)$ by $W^{-1}\oplus \sigma^{-1} I$ one gets $\Phi(\chi)$ as in \eqref{eq:Qchi}. This ends the proof.
\end{IEEEproof}
\section{Numerical Example}
\label{sec:NumExample}
\begin{figure}[!ht]
\centering
\psfrag{x}[][][1]{$\chi$}
\includegraphics[trim=2cm 0cm 1cm 0cm, clip, width=1\columnwidth]{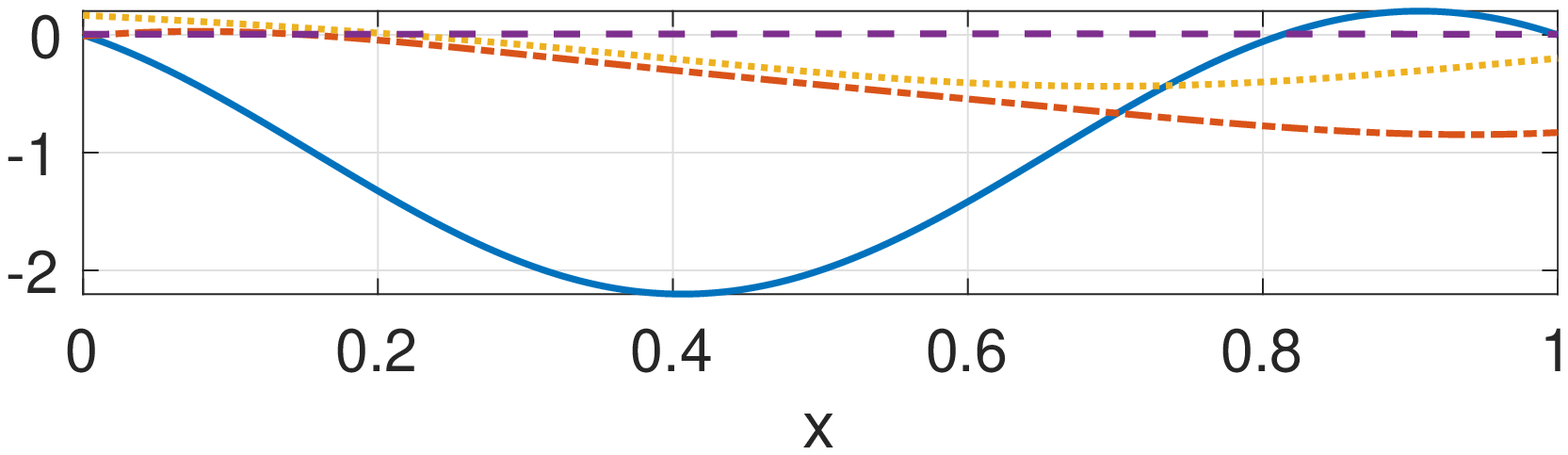}
\includegraphics[trim=2cm 0cm 1cm 0cm, clip, width=1\columnwidth]{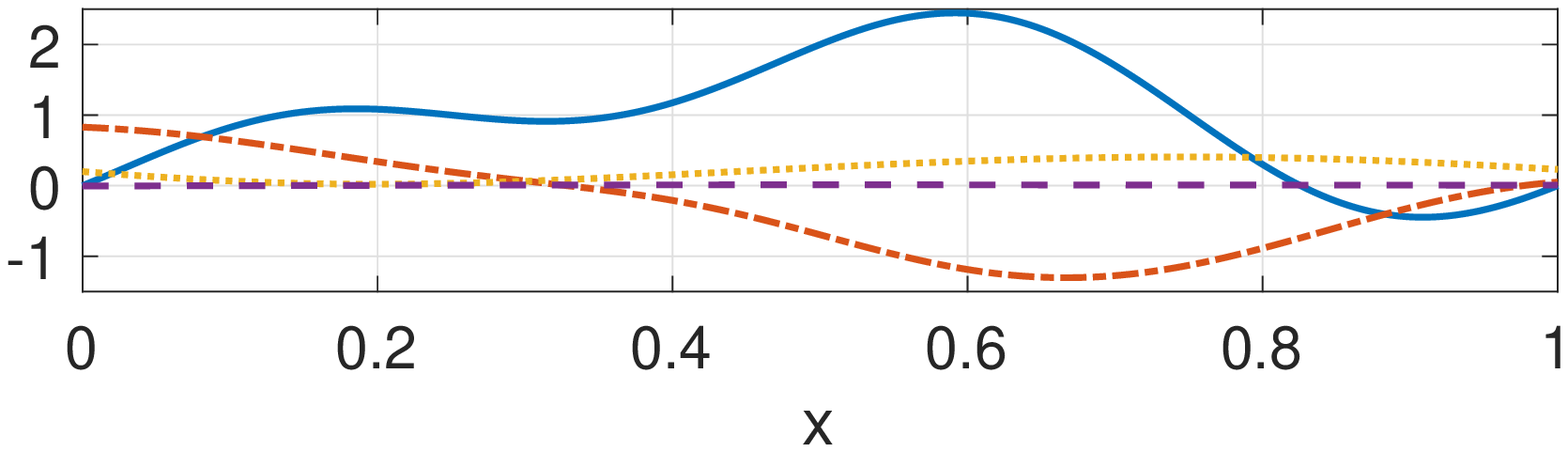}
 \caption{First (top) and second (bottom) components of $\overline{e}_s$ (defined in \eqref{eq:e_s}) for $t=0$ (solid blue), $t=2$ (dotted orange), $t=5$ (dash-dotted yellow), $t=20$ (dashed purple).}
 \label{fig:average}
\end{figure}

\begin{figure}[!ht]
\centering
\psfrag{x}[][][1]{$t$}
\includegraphics[trim=2cm 0cm 1cm 0cm, clip,  width=\columnwidth]{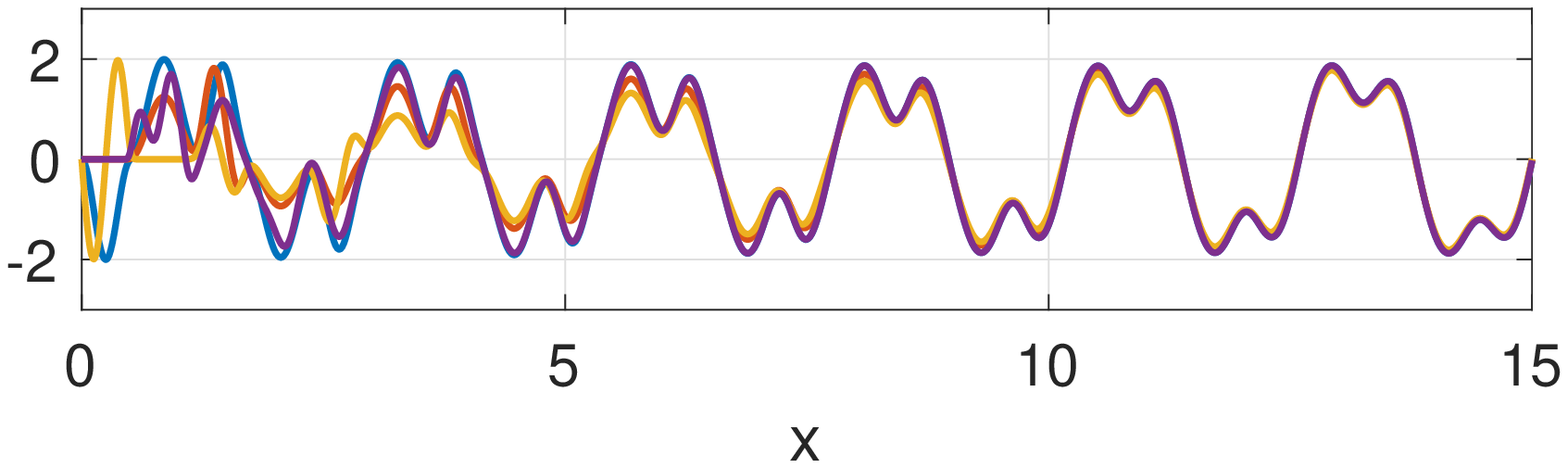}
\includegraphics[trim=2cm 0cm 1cm 0cm, clip, width=\columnwidth]{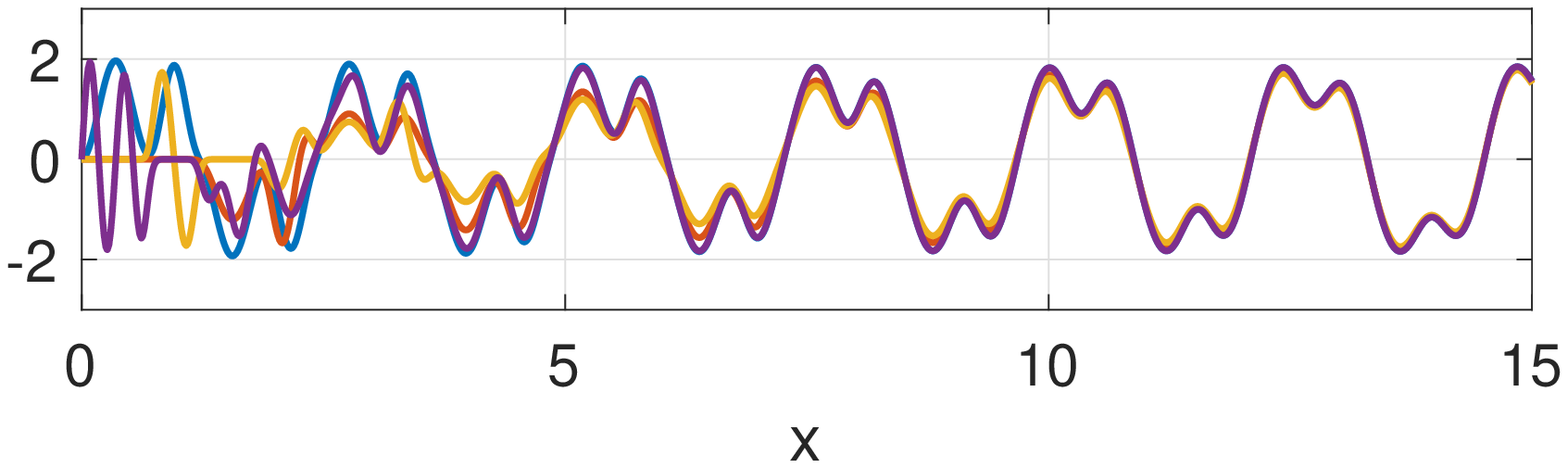}
 \caption{Time evolution of the first (top) and second (bottom) component of the boundary value $x_i(t, 1)$: $i=1$ (blue), $i=2$ (orange), $i=3$ (yellow), $i=4$ (purple).}
 \label{fig:boundary}
\end{figure}

In this example, we illustrate the effectiveness of the methodology illustrated
in Section~\ref{sec:Hyperbolic}. For this purpose, we consider a family of $n=4$, $2\times 2$ semilinear hyperbolic systems as in \eqref{eq:hyperbolic} with: $S=\begin{bmatrix}
2&0\\
0&\sqrt{2}
\end{bmatrix},E=\begin{bmatrix}
-0.1\\
0
\end{bmatrix},B=\begin{bmatrix}
2\\
0
\end{bmatrix}$, $H=\begin{bmatrix}0 & 1\\ -1 & 0\end{bmatrix}, Q=\begin{bmatrix}1 & 0 \end{bmatrix}, h(z)=\tanh(z)$.
Since the slope of $h$ belongs to $[0, 1]$, it is easy to show that $h$ satisfies \eqref{eq:QuadraticHyperbolic} with $G_{12}=1$ and $G_{22}=-2$. The agents are connected according to \eqref{eq:controlLap} with the Laplacian matrix $L=\left[\begin{smallmatrix} 
0&0&0&0\\ -1&2&-1&0\\0&-1&1&0\\-1&0&0&1
\end{smallmatrix}\right]$.
To design $K$, we solve the conditions 
in Corollary~\ref{coro:design} by performing a linear search on the scalar $\mu$, this leads to\footnote{Numerical solutions to SDP problems are obtained in Matlab\textsuperscript{\tiny\textregistered} via \emph{SDPT3} \cite{tutuncu2003solving} thanks to \emph{YALMIP} \cite{lofberg2004yalmip}. Numerical integration of hyperbolic PDEs is performed via the use of the Lax-Friedrichs (Shampine's two-step variant) scheme implemented in Matlab\textsuperscript{\tiny\textregistered} by Shampine \cite{shampine2005solving}.
Code at \scalebox{0.8}{\url{https://github.com/f-ferrante/FerranteCasadeiPrieurLCSS2021}}.}
$\mu=0.1474, W=\left[\begin{smallmatrix}11.799 & 0\\ 0 & 16.273 \end{smallmatrix}\right], K=\left[\begin{smallmatrix}0&-0.309 \end{smallmatrix}\right], \sigma=79.6164$. \figurename~\ref{fig:average} shows the evolution of the spatial average synchronization error, i.e.,
\begin{equation}
\label{eq:e_s}  
\chi\mapsto \overline{e_s}(\chi)\coloneqq\frac{1}{3}\left(\sum_{i=2}^4x_1(t, \chi)-x_i(t, \chi)\right),
\end{equation} 
for several values of $t$ from the initial condition:
$x_1(0, \chi)=(\cos(2\pi\chi)-1, 1-\cos(2\pi\chi))\, ,
x_2(0, \chi)=(0, 0)\, ,
x_3(0, \chi)=(2\sin(2\pi\chi), 0)\, ,
x_4(0, \chi)=(0, -2\sin(4\pi\chi))$. The picture clearly shows that the average synchronization error approaches zero, thereby ensuring asymptotic synchronization of the network. 
To further emphasize the effectiveness of the proposed methodology to enforce synchronization even in the presence of nontrivial regimes, in \figurename~\ref{fig:boundary} we report the time evolution of the boundary values $x_i(t, 1)$'s. The figure suggests that the considered network converges towards a nonvanishing trajectory.   
\section{Conclusion}
The problem of synchronizing a family of infinite-dimensional semilinear boundary actuated systems has been considered. Sufficient conditions in terms of operator inequalities have been proposed to ensure asymptotic synchronization.   
When specialized to the case of semilinear hyperbolic PDEs, these conditions  can be recast into a collection of matrix inequalities that can be easily checked. Thus, sufficient conditions for the design of the coupling gain are given in terms of linear matrix inequalities. The extension of the proposed results to different classes of infinite-dimensional systems as well as to heterogeneous networks are currently part of our ongoing research. 
\bibliography{Biblio}  

\appendix
In this section, we deal with Hilbert space-valued functions for which the following notations are used. Let $\mathscr{U}\subset\R$ be an interval and $\mathscr{V}$ be a separable real Hilbert space, $\mathcal{L}_1(\mathscr{U}; \mathscr{V})$ denotes the collection of equivalence classes of (strongly) measurable functions $f\colon \mathscr{U}\rightarrow \mathscr{V}$ such that $\int_\mathscr{U} \vert f(x)\vert_{\mathscr{V}} dx$ is finite. Similarly, we say that $f\in\mathcal{L}_1^{\textrm{loc}}(\mathscr{U}; \mathscr{V})$ if for any compact set $\Omega\subset\mathscr{U}$, $\left. f\right\vert_\Omega\in\mathcal{L}_1(\Omega; \mathscr{V})$. Moreover, in the remainder of the paper, the symbol $\nats$ denotes the set of positive integers.
\begin{IEEEproof}[Proof of Proposition~\ref{prop:existence}]
As a first step, we show that under the considered assumptions,  ``integrated solutions'' as in Definition~\ref{def:solution} and solutions to the integral equation \eqref{eq:SemigroupRep} coincide. From \cite[page 218]{magal2018theory}, since $\dom\Aop$ is dense in $\calX$ ($\Aop$ is the infinitesimal generator of a strongly continuous semigroup on $\calX$) it follows that  any ``integrated solution'' as in Definition~\ref{def:solution} satisfies 
\eqref{eq:SemigroupRep}. Now we show the other implication. 
Let $\varphi$ be a solution to \eqref{eq:SemigroupRep} with $x_0\in\calX$. Notice that since $\Aop$ is the infinitesimal generator of a strongly continuous semigroup on $\calX$ and $f$ is globally Lipschitz continuous,
from \cite[Theorem 1.2., page 185]{pazy2012semigroups} it follows that for all $x_0\in\calX$, there exists unique solution to \eqref{eq:SemigroupRep}. For the sake of exposition, let us rewrite $\varphi$ as follows:
\begin{equation}
\label{eq:solSemi}
\varphi(t)=\mathscr{T}(t)x_0+\Psi(t)\qquad\forall t\geq 0
\end{equation}
where:
$$
\Psi(t)\coloneqq\int_{0}^t \mathscr{T}(t-\theta)f(\varphi(\theta))d\theta
$$
Observe that $\varphi$ is continuous. Moreover, since $\Aop$ is the infinitesimal generator of the strongly continuous semigroup $\{\mathscr{T}(t)\}_{t\in\R_{\geq 0}}$, from 
\cite[Theorem 2.1.10, page 21]{curtain1995introduction} it follows that for all $t\in\R_{\geq 0}$
\begin{equation}
\label{eq:SemDom}
\begin{aligned}
&\int_{0}^t\mathscr{T}(s)x_0 ds\in\dom\Aop,&\Aop\int_{0}^t\mathscr{T}(s)x_0 ds=\mathscr{T}(t)x_0-x_0
\end{aligned}
\end{equation}
Pick any $t\in\R_{\geq 0}$, then one has:
$$
\int_0^t\Psi(s)ds=\int_{0}^t \int_{0}^s \mathscr{T}(s-\theta)f(\varphi(\theta))d\theta ds
$$
Hence, due to continuity of 
$s\mapsto\mathscr{T}(s-\cdot)f(\varphi(\cdot))$ and $\theta\mapsto\mathscr{T}(\cdot-\theta)f(\varphi(\theta))$, thanks to Fubini's theorem and by performing a simple change of variables, one gets:
\begin{equation}
\label{eq:Fubini}
\int_0^t\Psi(s)ds=\int_{0}^t \underbrace{\left(\int_{0}^{t-\theta} \mathscr{T}(\rho)f(\varphi(\theta))d\rho\right)}_{g_t(\theta)} d\theta
\end{equation}
At this stage notice that since $\Aop$ is the infinitesimal generator of the strongly continuous semigroup $\{\mathscr{T}(t)\}_{t\in\R_{\geq 0}}$, from 
\cite[Theorem 2.1.10, page 21]{curtain1995introduction} it follows that for all $\theta\in[0, t]$, $g_t(\theta)\in\dom\Aop$ and in particular 
\begin{equation}
\Aop g_t(\theta)=\mathscr{T}(t-\theta)f(\varphi(\theta))-f(\varphi(\theta))\qquad\forall \theta\in[0, t]
\label{eq:IntegralAg}
\end{equation}
Moreover, by observing that, due to continuity of $f$, for all $\theta\in[0, t]$, $g_t\in\mathcal{L}_1((0, \theta); \mathcal{X})$ and, thanks to \eqref{eq:IntegralAg}, $\Aop g_t\in\mathcal{L}_1((0, \theta); \mathcal{X})$. Thus, since $\Aop$ is closed ($\Aop$ is the infinitesimal generator of a strongly continuous semigroup; see \cite[Theorem 2.1.10, page 21]{curtain1995introduction}) by applying \cite[Theorem 3.7.12]{hille1996functional} it follows that \begin{equation}
\label{eq:SemDom2}
\begin{aligned}
\int_0^t\Psi(s)ds\in\dom\Aop,\,\,\, &\Aop\int_0^t\Psi(s)ds=\int_0^t\Aop\Psi(s)ds=\\
&\int_0^t\Aop g_t(\theta)d\theta
\end{aligned}
\end{equation}
where the second equality comes from \eqref{eq:Fubini}.
Therefore, from \eqref{eq:SemDom} and \eqref{eq:SemDom2} it follows that for all $t\geq 0$
$$
\int_0^t\varphi(s)ds\in\dom\Aop
$$
Now we show that $\varphi$ satisfies the identify in Definition~\ref{def:solution}. Let $t\geq 0$, from \eqref{eq:SemDom2}, by using \eqref{eq:IntegralAg} one gets
$$
\Aop\int_0^t\Psi(s)ds=\int_0^t (\mathscr{T}(t-\theta)f(\varphi(\theta))-f(\varphi(\theta)))d\theta
$$
Hence, by recalling \eqref{eq:solSemi} and combining the above expression with \eqref{eq:SemDom} one has:
$$
\Aop\int_0^t\!\!\!\varphi(s)ds=\mathscr{T}(t)x_0-x_0+\int_0^t (\mathscr{T}(t-\theta)f(\varphi(\theta))-f(\varphi(\theta)))d\theta
$$
Therefore using the expression of $\varphi$ in \eqref{eq:SemigroupRep}, one has
$$
\varphi(t)-\Aop\int_0^t \varphi(s)ds=x_0+\int_0^t f(\varphi(\theta))d\theta
$$
which, recalling that $\varphi(0)=x_0$, corresponds to the identity in Definition~\ref{def:solution}. The steps carried out so far enable to fully replace the notion of solution in Definition~\ref{def:solution} with \eqref{eq:SemigroupRep}. In particular, due to the above mentioned uniqueness of solutions  to \eqref{eq:SemigroupRep}, existence, uniqueness, and completeness of maximal solutions to \eqref{eq:AbstractODE} is established.

To conclude the proof, notice that since $\calX$ is a Hilbert space (an so it is reflexive) and $f$ is Lipschitz continuous, if $x_0\in\dom\Aop$, from \cite[Theorem 1.6., page 189]{pazy2012semigroups} the unique maximal solution to \eqref{eq:AbstractODE} from $x_0$ is a strong solution. 
\end{IEEEproof}
\medskip

\begin{IEEEproof}[Proof of Theorem~\ref{thm:GES}]
First we prove global exponential stability of $\A$ for strong solutions to \eqref{eq:AbstractODE}; see Proposition~\ref{prop:existence}.
In particular, pick $x_0\in\dom\Aop$ and let $\varphi$ be the unique maximal solution to \eqref{eq:AbstractODE} 
with $\varphi(0)=x_0$. From Proposition~\ref{prop:existence}, it follows that $\varphi$ is differentiable for almost all $t\geq0$, $\dot{\varphi}\in\mathcal{L}_1^{\rm{loc}}(\R_{\geq 0}; \calX)$, and
$$
\dot{\varphi}(t)=\Aop\varphi(t)+f(\varphi(t))\qquad\text{for\,\,almost\,\,all}\,\,t\in\R_{\geq 0}
$$
In particular, notice that since strong solutions are solutions in the sense of Definition~\ref{def:solution}, it turns out that for all $t\geq 0$:
\begin{equation}
\label{eq:AbsCont}
\begin{aligned}
\varphi(t)=&\varphi(0)+\Aop\int_0^t\varphi(s)ds+\int_0^tf(\varphi(s))ds\\
&=\varphi(0)+\int_0^t\Aop\varphi(s)ds+\int_0^t f(\varphi(s))ds\\
&=\varphi(0)+\int_0^t \dot{\varphi}(s)ds
\end{aligned}
\end{equation}
where the first identify follows from \cite[Theorem 3.7.12]{hille1996functional} due to $\Aop$ being closed, $\varphi(t)\in\dom\Aop$ for almost all $t\geq 0$, and $\Aop\varphi\in\mathcal{L}_1^{\rm loc}(\R_{\geq 0};\calX)$.
Namely, \eqref{eq:AbsCont} shows that $\varphi$ is locally absolutely continuous on $\R_{\geq 0}$.
Define for all $t\geq 0$, $W(t)\coloneqq V(\varphi(t))$.
Observe that since $V$ is Fréchet differentiable everywhere and $\varphi$ is differentiable almost everywhere, it follows that for almost all $t\geq 0$.
$$
\dot{W}(t)\coloneqq \frac{d}{dt}W(\varphi(t))=DV(\varphi(t))\dot{\varphi}(t) 
$$
In particular, since $V$ is Fréchet differentiable (and so locally Lipschitz), $\phi$ is continuous, and $\dot{\varphi}\in\mathcal{L}_1^{\rm{loc}}(\R_{\geq 0}; \calX)$, if follows that $W=V\circ\varphi$ is locally absolutely continuous on $\R_{\geq 0}$. Then, since $\varphi$ is a strong solution to \eqref{eq:AbstractODE}, for all almost all $t\geq 0$ one has:
$$
\dot{W}(t)=DV(\varphi(t))(\Aop\varphi(t)+f(\varphi(t)))
$$
Hence, using items $(i)$ and $(ii)$, for almost all $t\geq 0$ one has:
$$
\dot{W}(t)\leq-\frac{\alpha_3}{\alpha_2}W(t)
$$
which, using the fact that $W$ is locally absolutely continuous on $\R_{\geq 0}$, 
Therefore, a direct application of the comparison lemma gives:
$$
W(t)\leq W(0)e^{-\frac{\alpha_3}{\alpha_2}t}\qquad\forall t\in\R_{\geq 0}
$$
which thanks to item $(i)$ gives:
$$
d(\varphi(t),\mathcal{S})\leq \left(\frac{\alpha_2}{\alpha_1}\right)^{\frac{1}{p}} e^{-\frac{\alpha_3}{p\alpha_2} t}d(\varphi(0),\mathcal{S})\qquad\forall t\in\R_{\geq 0}
$$
thereby showing global exponential stability of $\A$ for strong solutions. We now extend the proof to all solutions to \eqref{eq:AbstractODE}. Let $\varphi$ be a maximal solution to \eqref{eq:AbstractODE}. Since, $\dom\Aop$ by assumption $\Aop$ is the infinitesimal generator of a $\mathcal{C}_0$ semigroup $\{\mathcal{T}(t)\}_{t\in\R_{\geq 0}}$ on $\calX$, from \cite{arendt2011vector}  it follows that $\dom\Aop$ is dense in $\calX$. Thus, there exists a sequence $\{x^k_0\}_k\subset\dom\Aop$ such that $x^k_0\rightarrow\varphi(0)$. Now, let for all $k\in\nats$
$$
\varphi_k(t)=\mathscr{T}(t)x^k_0+\int_{0}^t\mathscr{T}(t-s)f(\varphi_k(s))\qquad\forall t\geq 0
$$
Observe that for all $k\in\nats$, thanks to Proposition~\ref{prop:existence}, $\varphi_k$ is a strong solution to \eqref{eq:AbstractODE}. Thus for all $k$, as shown earlier:
\begin{equation}
d(\varphi_k(t),\mathcal{S})\leq \left(\frac{\alpha_2}{\alpha_1}\right)^{\frac{1}{p}} e^{-\frac{\alpha_3}{p\alpha_2} t}d(\varphi_k(0), \mathcal{S})\qquad\forall t\in\R_{\geq 0}
\label{eq:BoundGESk}
\end{equation}
Moreover, observe that thanks to \eqref{eq:SemigroupRep}, for all $t\in\R_{\geq 0}$
\begin{equation}
\lim_{k\rightarrow \infty}\vert\varphi_k(t)-\varphi(t)\vert_\calX=0
\label{eq:NormConv}
\end{equation}
which, by using the following elementary inequality\footnote{We use the following property: Let $X$ be a normed linear vector space and $A\subset X$ be nonempty. Then, for all $x, y\in X$ one has
$$
d(x, A)\leq \Vert x-y\Vert_X+d(y, A)
$$}
$$
\vert d(\varphi(t),\mathcal{S})-d(\varphi_k(t),\mathcal{S})\vert\leq \vert  \varphi(t)- \varphi_k(t)\vert_\calX,\,\,\,\forall t\geq 0, k\in\nats
$$
yields for all $t\geq0$
$$
\lim_{k\rightarrow\infty}d(\varphi_k(t),\mathcal{S})=d(\varphi(t), \mathcal{S})
$$
Therefore, taking the limit on $k$ in \eqref{eq:BoundGESk} and using the above relationship establish the result.
\end{IEEEproof}
\begin{lemma}
\label{appendix:lemmino}
Consider the matrix $M$ defined in \eqref{eq:Tcv} and  let $\mathcal{K}\in\mathscr{L}(\mathcal{Q}\oplus\mathcal{E})$ be selfadjoint such that \eqref{eq:QadraticIneq} is satisfied. Define:
\begin{eqnarray}
\mathfrak{M} &\coloneqq& \begin{bmatrix}
(M^\top \otimes I_\mathcal{\calX}) & 0\\
0 & (M^\top \otimes I_{\mathfrak{E}})\end{bmatrix}\\
\mathfrak{K} &\coloneqq&\left[
\begin{array}{c|c}
   I_{n-1} \otimes \mathfrak{Q}^\star \mathcal{K}_{11} \mathfrak{Q} & I_{n-1} \otimes  \mathfrak{Q}^\star \mathcal{K}_{12}  \\[.6em]
  \hline
 I_{n-1} \otimes \mathcal{K}_{12}^\star \mathfrak{Q}  & I_{n-1} \otimes \mathcal{K}_{22} \\
\end{array}
\right].
\end{eqnarray} 
Then, for all $\vartheta \in\mathcal{X}^{n-1}$
\begin{equation}\label{eq:lemma2}
\begin{array}{l}
\left\langle \mathfrak{M}\begin{bmatrix}
\vartheta \\
F((I_n \otimes \mathfrak{Q}) \vartheta)
\end{bmatrix}, \mathfrak{K}\mathfrak{M}\begin{bmatrix}
\vartheta \\
F((I_n \otimes \mathfrak{Q}) \vartheta)
\end{bmatrix}\right\rangle =\\ \hspace*{0cm} \displaystyle \frac{1}{n}\sum_{1 \leq i \leq j \leq n} \left\langle 
\begin{bmatrix}
\mathfrak{Q} \vartheta_i- \mathfrak{Q}\vartheta_j\\
f(\mathfrak{Q} \vartheta_i)-f(\mathfrak{Q} \vartheta_j)
\end{bmatrix},\mathcal{K}\begin{bmatrix}
\mathfrak{Q} \vartheta_i-\mathfrak{Q} \vartheta_j\\
f(\mathfrak{Q} \vartheta_i)-f(\mathfrak{Q} \vartheta_j)
\end{bmatrix}\right\rangle
\end{array}
\end{equation}
\end{lemma}
\begin{IEEEproof}
Using the Kronecker product introduced in Definition~\ref{def:kron}, one has 
$$
\mathfrak{M}\mathfrak{K}=\mathfrak{K}\mathfrak{M}
$$ 
This in turn, by relying on Lemma~\ref{lemm:kron}, gives
\begin{align*}
\label{eq:frakMbar}
\left\langle\mathfrak{M} \begin{bmatrix}
\vartheta\\
F((I_n \otimes \mathfrak{Q}) \vartheta)
\end{bmatrix}, \mathfrak{K} \mathfrak{M} \begin{bmatrix}
\vartheta\\
F((I_n \otimes \mathfrak{Q}) \vartheta)
\end{bmatrix} \right\rangle=\\
\left\langle\overline{\mathfrak{M}} \begin{bmatrix}
\zeta\\
F(\zeta)
\end{bmatrix}, \overline{\mathfrak{K}}\begin{bmatrix}
\zeta\\
F(\zeta)
\end{bmatrix} \right\rangle
\end{align*}
where $\zeta\coloneqq (I_n \otimes \mathfrak{Q}) \vartheta$, 
\begin{align}
\overline{\mathfrak{K}}& \coloneqq\begin{bmatrix}
I_{n-1} \otimes \mathcal{K}_{11} & I_{n-1} \otimes \mathcal{K}_{12}\\
I_{n-1} \otimes \mathcal{K}_{12}^\star & I_{n-1} \otimes \mathcal{K}_{22}
\end{bmatrix}\\
\overline{\mathfrak{M}}& \coloneqq\begin{bmatrix}
(MM^\top \otimes I_\calX) & 0\\
0 & (MM^\top \otimes I_{\mathfrak{E}})\end{bmatrix}
\end{align}
The latter, similarly as in \cite[Lemma 1]{zhang2014fully}, gives
\[
\begin{aligned}
&\left\langle\overline{\mathfrak{M}} \begin{bmatrix}
\zeta\\
F(\zeta)
\end{bmatrix}, \overline{\mathfrak{K}}\begin{bmatrix}
\zeta \\
F(\zeta)
\end{bmatrix} \right\rangle=\\
&\frac{1}{n} \sum_{1 \leq i \leq j \leq n} \left\langle
\begin{bmatrix}
\mathfrak{Q}\vartheta_i-\mathfrak{Q}\vartheta_j\\
f(\mathfrak{Q} \vartheta_i)-f(\mathfrak{Q} \vartheta_j)
\end{bmatrix},\mathcal{K}
\begin{bmatrix}
\mathfrak{Q}\vartheta_i-\mathfrak{Q}\vartheta_j\\
f(\mathfrak{Q}\vartheta_i)-f(\mathfrak{Q} \vartheta_j)
\end{bmatrix}
\right\rangle
\end{aligned}
\]
which, by using \eqref{eq:frakMbar}, yields \eqref{eq:lemma2}.
\end{IEEEproof}
\medskip
\begin{lemma}
\label{lemm:kron}
Let $S\in\R^{\theta\times\theta}$, $(V, \langle \cdot, \cdot\rangle_V)$ and $(U, \langle \cdot, \cdot\rangle_U)$ be real Hilbert spaces, and $\Pi\in\mathscr{L}(U,V)$. Define $(V^\theta,  \langle\cdot, \cdot\rangle_{V^\theta})$, $(U^\theta,  \langle\cdot, \cdot\rangle_{U^\theta})$ and let $(S\otimes \Pi)\in\mathscr{L}(U^\theta, V^\theta)$. Then, the following identity holds:
\begin{equation}
\label{eq:adjointKron}
(S\otimes \Pi)^\star=S^\top\otimes\Pi^\star
\end{equation}
\end{lemma}
\begin{IEEEproof}
For the sake of notation, we assume that $U=V$, the extension of the proof to the general case is trivial.
Let $\mathcal{H}_1\coloneqq (\R^{\theta}\tens V, \langle\cdot, \cdot\rangle_{\mathcal{H}_1})$ 
where $\R^{\theta}\tens V$ is the tensor product of $\R^\theta$ and $V$ 
and:
$$
\begin{aligned}
\langle x\tens y, z\tens w\rangle_{\mathcal{H}_1}&\coloneqq \langle x, z\rangle_{\R^\theta} \langle y, w\rangle_{V}\\
&\quad\quad\quad\quad\forall x\tens y, z\tens w\in \mathcal{H}_1
\end{aligned}
$$
In particular, we identify $\R^{\theta}\tens V$ with $V^\theta$ with the following tensor product:
$$
x\tens y\coloneqq \begin{bmatrix}
x_1 y\\
x_2 y\\
\vdots\\
x_\theta y
\end{bmatrix}\quad \forall (x, y)\in\R^\theta\times V
$$
With that, for all $x\tens y\in\mathcal{H}_1$
$$
(S\tens \Pi)(x\tens y)\coloneqq(Sx)\tens(\Pi y)=(S\otimes \Pi)(x\tens y)
$$
that is, $S\tens \Pi$ is identified with $S\otimes \Pi$ on $\mathcal{H}_1$.
At this stage, notice that for all $x_1\tens y_1, x_2\tens y_2\in\mathcal{H}_1$:
\begin{equation}
\begin{aligned}
\left\langle (S\tens \Pi)(x_1\tens y_1), x_2\tens y_2
\right\rangle_{\mathcal{H}_1}&=\langle Sx_1, x_2\rangle_{\R^\theta} \langle \Pi y_1, y_2\rangle_{V}\\
&=\langle x_1, S^\top x_2\rangle_{\R^\theta} \langle y_1, \Pi^\star y_2\rangle_{V}\\
&=\langle x_1\tens y_1, (S^\top\tens\Pi^\star)(x_2\tens y_2)
\rangle_{\mathcal{H}_1}
\end{aligned}
\label{eq:FundamentalInner}
\end{equation}
This shows the desired identity on $\mathcal{H}_1$. To end the proof, we show that $\mathcal{H}_1$ and $\mathcal{H}_2\coloneqq(V^\theta,  \langle\cdot, \cdot\rangle_{V^\theta})$ are isometrically isomorphic through the inclusion mapping $\iota\colon \mathcal{H}_1\rightarrow\mathcal{H}_2$. In particular, for all $h=x\tens v\in\mathcal{H}_1$
$$
\langle  \iota h,  \iota h\rangle_{\mathcal{H}_2}=\sum_{i=1}^\theta \langle x_i v, x_i v\rangle_{V}=\Vert v\Vert_V^2\sum_{i=1}^\theta x_i^2
$$
On the other hand
$$
\langle h, h\rangle_{\mathcal{H}_1}=\langle x, x\rangle_{\R^\theta}\langle v, v\rangle_{V}=\langle  \iota h,  \iota h\rangle_{\mathcal{H}_2}
$$
this shows that $\mathcal{H}_1$ and $\mathcal{H}_2$ are isometrically isomorphic, namely $\iota$ is unitary, i.e., $\iota^{-1}=\iota^\star$. For all 
$h_1, h_2\in\mathcal{H}_2$
$$
\begin{aligned}
&\left\langle (S\otimes \Pi)\iota^\star h_1, \iota^\star h_2
\right\rangle_{\mathcal{H}_1}=
\left\langle \iota (S\otimes \Pi)\iota^\star h_1,  h_2
\right\rangle_{\mathcal{H}_2}
\end{aligned}
$$
$$
\begin{aligned}
\left\langle \iota^\star h_1, (S^\top\otimes\Pi^\star)\iota^\star h_2
\right\rangle_{\mathcal{H}_1}&=\left\langle h_1, \iota (S^\top\otimes\Pi^\star)\iota^\star h_2
\right\rangle_{\mathcal{H}_2}
\end{aligned}
$$
which from \eqref{eq:FundamentalInner} yields:
$$
\left\langle \iota (S\otimes \Pi)\iota^\star h_1,  h_2
\right\rangle_{\mathcal{H}_2}=
\left\langle h_1, \iota (S^\top\otimes\Pi^\star)\iota^\star h_2\right\rangle_{\mathcal{H}_2}
$$
where $\iota (S^\top\otimes\Pi^\star)\iota^\star=(S^\top\otimes\Pi^\star)\in\mathscr{L}(\mathcal{H}_2)$, 
that is \eqref{eq:adjointKron}.
\end{IEEEproof}
\end{document}